\documentclass[10pt,conference]{IEEEtran}
\IEEEoverridecommandlockouts
\usepackage{cite}
\usepackage{amsmath,amssymb,amsfonts}
\usepackage{algorithmic}
\usepackage{graphicx}
\usepackage{textcomp}
\usepackage{xcolor}
\usepackage{xspace}
\usepackage{subfigure}
\usepackage{wrapfig}
\usepackage{booktabs}
\usepackage{diagbox}
\def\BibTeX{{\rm B\kern-.05em{\sc i\kern-.025em b}\kern-.08em
    T\kern-.1667em\lower.7ex\hbox{E}\kern-.125emX}}

\newcommand{\method}{APC\xspace}

\begin{document}

\title{Adaptive Punishment for Cooperation in Mixed-Motive Games}



\author{
    Min Tang$^{1,2}$, 
    Fanqi Kong$^{2}$, 
    Linyuan Lü$^{1,\dagger}$, 
    Xue Feng$^{2,\dagger}$, \\

    $^1$University of Science and Technology of China, Hefei, China\\
    $^2$State Key Laboratory of General Artificial Intelligence, BIGAI, Beijing, China\\
    tangmin8554@mail.ustc.edu.cn, linyuan.lv@ustc.edu.cn
}

\maketitle

\renewcommand{\thefootnote}{\fnsymbol{footnote}} 
\footnotetext[2]{Corresponding author} 
\renewcommand{\thefootnote}{\arabic{footnote}} 

\begin{abstract}
Mixed-motive scenarios are ubiquitous in real-world multi-agent interactions, where self-interested agents often defect for immediate rewards, overlooking the potential of altruistic cooperation to improve long-term gains and collective welfare. Peer punishment can deter defection, but as costly second-order altruism, its persistent imposition may undermine the punisher’s interests. Existing approaches often struggle to effectively implement punishment to promote cooperation. To balance the efficacy and cost of punishment, we propose Adaptive Punishment for Cooperation (APC), a distributed method that determines punishment intensity based on both a dynamic punishment probability and the severity of defection. This dynamic probability substantially reduces costly and ineffective punishment while also promotes cooperation. To accurately assess defection and its severity, we use a defection awareness module, whose learning is guided by game reward. Theoretical analysis and empirical results show APC performs effectively in iterated public goods game. Empirically, APC also significantly outperforms existing baselines across sequential social dilemmas, learning rational and effective punishment policies that foster cooperation by strategically deterring defection.
\end{abstract}

\begin{IEEEkeywords}
decentralized multi-agent system, reinforcement learning, social dilemmas
\end{IEEEkeywords}

\section{Introduction}

Multi-Agent Reinforcement Learning (MARL) allows agents to interact in shared environments, yet mixed-motive scenarios \cite{action1,Adasociety} pose unique challenges compared to purely cooperative or competitive tasks \cite{MAddpg}. In these settings, dynamic agent relationships often lead to social dilemmas where short-term individual gains supersede long-term collective welfare. Traditional training paradigms often fail to resolve these issues: Centralized Training with Decentralized Execution (CTDE) \cite{VDN} relies on unrealistic assumptions of joint optimization among self-interested agents, while Decentralized Training (DTDE) frequently converges to suboptimal local equilibria (see Fig. \ref{curve}). Consequently, there is a critical need for methods that guide decentralized agents out of these dilemmas toward socially beneficial outcomes.

To promote cooperation in mixed-motive games, it is crucial to both deter defection and incentivize collective welfare. Inspired by human societies, punishment raises the cost of defection, discouraging short-term selfish behavior and guiding agents toward long-term group benefits \cite{punish1}. However, peer punishment imposes a penalty on others while incurring a cost for the punisher (e.g., time and effort spent stopping someone from smoking in public). That is, punishment is second-order altruism. Excessive punishment may harm the punisher’s own interest, and agents may avoid punishing due to its cost, failing to achieve cooperation \cite{assum3}. To address these challenges, we propose a distributed decision-making method named \method (Adaptive Punishment for Cooperation), which is context-sensitive and adaptively adjusts the probability and severity of punishment. Guided by reward signals, \method learns a defection awareness module to evaluate the defectiveness of others' actions, which determines the probability and intensity of punishment. Notably, to prevent ineffective punishment and avoidable cost, punishment probability is dynamically adjusted according to the historical effectiveness of punishment, based on the reduction in defections.

The main contributions of this work are:
\textbf{(1)} We propose a novel opponent-adaptive punishment method, APC, that dynamically adjusts punishment intensity to avoid ineffective punishment and excessive costs, thereby promoting cooperation in mixed-motive games.
\textbf{(2)} We develop a self-learning defection awareness algorithm for detecting different degrees of defection in multi-agent interactions.
\textbf{(3)} We provide theoretical analysis and demonstrate empirically that APC outperforms existing baselines across Iterated Public Goods Game and Sequential Social Dilemmas (SSDs).


\section{Related Work}

Multi-agent reinforcement learning has been widely applied to address problems related to social dilemmas. Various reward mechanisms, including centralized redistribution  \cite{DBLP:conf/atal/GempM0DBBT22} and decentralized peer-to-peer rewarding  \cite{gifting1}, have been proposed to promote cooperation by aligning individual and collective interests in mixed-motive games. Punishment, which promote cooperation without requiring additional reward resources, have gained increasing attention. In evolutionary game theory, punishment is widely regarded as a key factor in achieving stable cooperation \cite{control1,oppo4}. Some approaches \cite{control3,control4} rely on centralized controllers or self-organized structures among individuals to implement punishment, further investigating how punishment influence the evolution and stability of cooperative behaviors. Other methods \cite{condition1,condition2,condition3} adopt conditional punishment policy, where the decision to punish depends on conditions such as the difference between an individual's own payoff and the average payoff of its neighbors, thereby improving the specificity and effectiveness of punishment.

In MARL, although some studies have attempted to incorporate punishment method into multi-agent systems, some designs remain relatively simple—typically adding a punishment action into the action space \cite{action1,action2,action3}. However, experimental results, shown as in Figure~\ref{fig:training_curve}, have shown that merely relying on such punishment action combined with standard Independent MARL methods often fails to promote cooperation in SSDs. Other approaches \cite{assum1,assum4} depend on auxiliary mechanisms or specific environmental assumptions to support the effectiveness of punishment, such as introducing reputation systems or assuming that punishment behaviors can generate positive incentives. Some methods further assume that punishment can directly yield positive rewards for the punisher \cite{rl_punish1, rl_punish2}, which is clearly unrealistic. We propose \method, which does not rely on environmental assumptions but instead identifies defection behaviors through a defection predictor network and applies targeted punishment. This guides agents to learn reasonable and effective punishment policy, thereby promoting cooperation more effectively.

\section{Preliminaries}
An $N$-player Partially Observable Markov Game (POMG) is defined as $\mathcal{M} = \langle N, \mathcal{S}, \{ \mathcal{O}^i \}, \{ \mathcal{A}^i \}, T, \{ R^i \} \rangle$. Here, $\mathcal{S}$ is the state space, while $\mathcal{O}^i$ and $\mathcal{A}^i$ denote the observation and action spaces for agent $i$, respectively. The state transition function $T: \mathcal{S} \times \mathcal{A}^1 \times \dots \times \mathcal{A}^N \times \mathcal{S} \to [0,1]$ defines the probability distribution over $\mathcal{S}$, representing the probability of transitioning to state $s'$ given the current state $s$ and joint action $\vec{a} = (a^1, \dots, a^N)$. Each agent $i$ follows a policy $\pi^i(a^i|o^i)$ to maximize its expected discounted return:
\begin{equation}
    V_i^{\pi^i}(s_0) = \mathbb{E}_{\vec{a}_t \sim \pi, s_{t+1} \sim T} \left[ \sum_{t=0}^{\infty} \gamma^t R^i(s_t, \vec{a}_t) \right],
\end{equation}
where $\gamma \in [0,1)$ is the discount factor. In our decentralized MARL framework, agents are trained using the Advantage Actor-Critic (A2C) method. The policy $\pi^i$ is parameterized by $\theta^i$, and the critic evaluates the actor via the TD-error: $\delta_t^i = R^i(s_t, \vec{a}_t) + \gamma V^{\pi_{\theta^i}}(\vec{o}_{t+1}) - V^{\pi_{\theta^i}}(\vec{o}_{t})$, where $\vec{o}_t$ represents the joint observation at timestep $t$.

\section{Methodology}
\begin{figure}[t]
    \centering
    \includegraphics[width=0.485\textwidth]{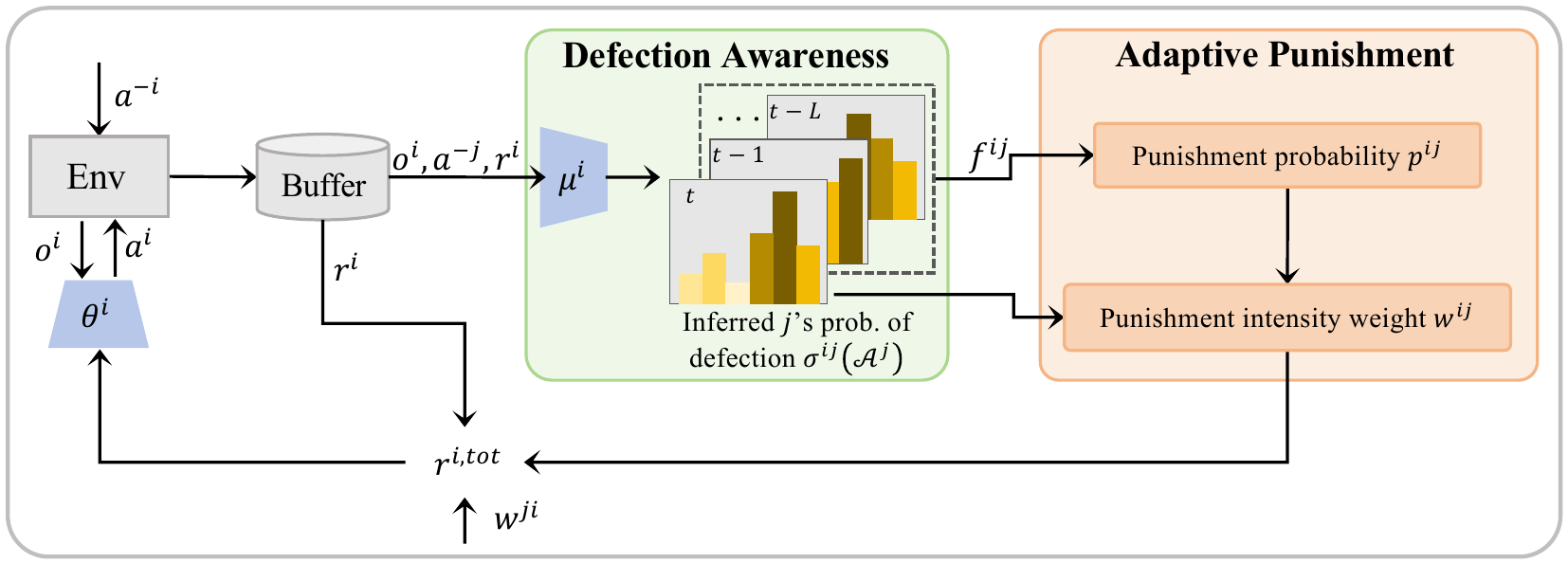}
    \caption{\textbf{Overview of the \method Framework.} 
    The framework consists of two core components: Defection Awareness and Adaptive Punishment. 
    Any Agent $i$ interacts with the environment to collect trajectories, which are stored in a replay buffer to train the defection predictor network $\mu^i$. 
    $\mu^i$ is trained using $o^i, a^{-j}$, and $r^i$ to predict the probability distribution $\sigma^{ij} = p_{\mu^i}(\cdot| o^i, a^{-j})$ over $\mathcal{A}^j$. 
    A higher-probability action $a^j$ indicates a greater degree of defection. 
    Adaptive punishment computes the probability $p^{ij}$ (Eq.(\ref{eq:punishment_prob})) based on the defection frequency $f^{ij}$ over the past $L$ timesteps. 
    The punishment intensity weight $w^{ij}$ is then determined: if $\sigma^{ij}(a^j) > \tfrac{1}{|\mathcal{A}^j|}$, $w^{ij}$ is set proportional to $\sigma^{ij}(a^j)$ with probability $p^{ij}$; otherwise, $w^{ij}=0$. 
    Finally, agent $i$’s total reward for policy update is $r^{i,\text{tot}} = r^i - \sum_{j=1, j \neq i}^{N} w^{ij}c - \sum_{j=1,j \neq i}^{N} w^{ji}\delta$, where $c$ and $\delta$ are unit cost and penalty.}
    \label{fig:dapc}
\end{figure}

To address the dilemma of cooperation and defection in multi-agent systems, we propose a distributed MARL method, named  Adaptive Punishment for Cooperation (\method). As shown in Fig.~\ref{fig:dapc}, \method consists of two core modules: defection awareness and adaptive punishment. The defection awareness module takes agents’ local observation trajectories as input and produces a probability distribution over the opponent’s actions, where actions with higher probabilities are regarded as more severe defections. The adaptive punishment module then uses both the defection judgment and the variation in defection frequency as input to generate a dynamic punishment intensity—that is, the more frequent the defections, the harsher the punishment. Moreover, it adjusts the punishment probability based on the effectiveness of punishment in reducing defections. Next, we take agent $i$ as the focal agent.

\subsection{Defection Awareness}
We propose a defection awareness method to train the Defection Predictor Network $\mu^i$ to identify defection, where the network learns to predict the behavior of a target agent $j$ that would harm agent $i$'s reward $r^i_t$ more. $\mu^i$ outputs probability distribution of defection $\sigma^{ij}_t = p_{\mu^i}(\cdot| o_t^i, a_t^{-j})$ over the agent $j$'s action space $\mathcal{A}^j$, given agent $i$’s observation $o_t^i$ and the joint actions of all other non-target agents $a_t^{-j}$. The dimensionality of $a_t^{-j}$ is fixed, and actions of agents outside of the observation range are represented using a default placeholder value of $-1$. If $
\sigma^{ij}_t(a_t^j) > \frac{1}{|\mathcal{A}^j |}$, $a_t^j$ is considered a defection by agent $i$. The training objective of $\mu^i$ is formalized as a maximization problem: 
\begin{equation}
\label{eq:instant_reward}
J(\mu^i) = \mathbb{E}_{a^j_t \sim p_{\mu^i}} \left[ -r^i_t(s_t, \vec{a}_t) + \beta H(\sigma^{ij}_t) \right].
\end{equation}

Here, $-r^i_t(s_t,\vec{a}_t)$ drives the network to minimize the reward, and $\beta H(\sigma^{ij}_t)$ incorporates entropy regularization to encourage behavioral diversity and prevent premature convergence to a peaked distribution. $H(\sigma^{ij}_t) = \mathbb{E}_{a^j_t \sim \sigma^{ij}_t} \big[ - \log \sigma^{ij}_t(a^j_t) \big]$ denotes the entropy of the distribution $\sigma^{ij}_t$, which measures its randomness: a larger entropy indicates a more uniform distribution. The regularization parameter $\beta$ the entropy's importance in the optimization objective. The objective function $J(\mu^i)$ is designed to guide agent $i$ toward learning to identify defection with sufficient exploration.  

Through this process, $\mu^i$ effectively identifies the behaviors of target agent $j$ that would more harm agent $i$’s reward, given different $o^i_t$ and $a_t^{-j}$. Once $\mu^i$ has converged, its parameters are fixed during the training of policy network $\theta^i$, ensuring stable defection awareness. This learned policy serves as a criterion for identifying defection actions in an opponent-aware manner. The entire training is performed via gradient descent using multi-agent trajectory data. 

\subsection{Adaptive Punishment}
We introduce an adaptive punishment mechanism to guide agents toward rational and graded punitive behavior. Adaptivity is reflected in two aspects: (1) punishment probability is dynamically adjusted according to whether past punishments have successfully reduced defections; and (2) the punishment intensity is modulated based on the degree of defection.

The punishment probability $p_t^{ij}$ denotes the probability with which agent $i$ punishes agent $j$ upon defection. $p_t^{ij}$ is dynamically adjusted to prevent ineffective punishment, in which past punishments have failed to reduce defections. It is evaluated by the change of defection frequency of agent $j$ from the perspective of agent $i$, denoted as $f_t^{ij}$, where defection is determined by $\mu_i$. Specifically, $f_t^{ij,s}$ denotes the value of $f_t^{ij}$ in window $s$, where each window spans $L$ timesteps. If $f_t^{ij,s}$ does not decrease compared to $f_t^{ij,s-1}$, the punishment in window $s$ is deemed ineffective. Consequently, $p_t^{ij}$ should be reduced in window $s+1$. We compute $p_t^{ij}$ in window $m$ as follows:
\begin{equation}
\label{eq:punishment_prob}
p_t^{ij} = 1 - \frac{1}{m-1} \sum_{s=1}^{m-1} \mathbf{1} \left[ \mathcal{C}_s \right],
\end{equation}
where $\mathcal{C}_s$ represents the condition for the punishment being considered ineffective in window $s$, defined as:
\begin{equation}
\label{eq:condition_detail}
\mathcal{C}_s = \left( \left( f_t^{ij,s} \geq f_t^{ij,s-1} \lor \left| f_t^{ij,s} - \bar{f}_t^{ij,s} \right| < \varepsilon \right) \land f_t^{ij,s} \geq \varepsilon \right),
\end{equation}
where $\bar{f}_t^{ij,s} = \frac{1}{s} \sum_{k=1}^{s} f_t^{ij,s-k}$ is the average defection frequency over the past $s$ windows. $\mathbf{1}[\cdot]$ is the indicator function that outputs 1 if the condition inside is true, indicating that the punishment in window $s$ is considered ineffective, and 0 otherwise and $\varepsilon$ is a tolerance threshold.  The condition $f_t^{ij,s} \geq f_t^{ij,s-1}$ indicates that the punishment in window $s$ failed to reduce $f_t^{ij}$. The term $\left| f_t^{ij,s} - \frac{1}{s} \sum_{k=1}^{s} f_t^{ij,s-k} \right| < \varepsilon$ suggests that $f_t^{ij}$ in window $s$ has not significantly decreased compared to the average defection frequency over the past $s-1$ windows. The constraint $f_t^{ij,s} \geq \varepsilon$ avoids misinterpreting a low defection frequency as signs of ineffectiveness. When $f_t^{ij,s}$ is very small, the punishment in window $s$ is still effective because it helps maintain cooperation. These suggest that agent $j$ is unlikely to change its behavior in response to punishment, and continuing to apply punishment in such cases would lead to unnecessary costs and a decrease in $r_t^{i,\text{tot}}$. Specifically, punishment is considered effective by default in window $0$ and $1$. Accordingly, $p_t^{ij}$ is initialized to 1 in window $0$ and $1$, and is subsequently updated after window $1$ based on Eq.(\ref{eq:punishment_prob}).

Any agent $i$ holds a Punishment Intensity Weight (PIW) vector at time step $t$: $w_t^i = \big[ w_t^{ij} \big]_{j=1}^N$, where $w_t^{ij} \in [0,1]$ is the fraction of agent $i$’s punishment to agent $j$. Let $B_t^{ij}$ be a Bernoulli random variable with success probability $p_t^{ij}$, i.e., $B_t^{ij} \sim \mathrm{Bernoulli}(p_t^{ij})$. $w_t^{ij}$ is computed as follows:
\setlength\abovedisplayskip{4pt} 
\setlength\belowdisplayskip{4pt} 
\begin{equation}
\label{eq:punishment_intensity}
w_t^{ij}
= B_t^{ij}\;\cdot\;
\begin{cases}
0, & \sigma_t^{ij}(a_t^j)\le \dfrac{1}{|\mathcal A^j|},\\
\dfrac{\sigma_t^{ij}(a_t^j)}{\max\limits_{a\in\mathcal A^j}\sigma_t^{ij}(a)}, & \sigma_t^{ij}(a_t^j)>\dfrac{1}{|\mathcal A^j|}.
\end{cases}
\end{equation}
If agent $i$ punish agent $j$, agent $i$ pays $w_t^{ij}c$ and agent $j$ loses $w_t^{ij}\delta$. In partially observable environments, agent $i$ can only punish $j$ if $j$ is within its observation range; otherwise, punishment is not applied. Agent $i$’s total reward is
$r_t^{i,\text{tot}}
= r_t^i - \sum_{j=1, j \neq i}^{N} w_t^{ij}c
- \sum_{j=1,j \neq i}^{N} w_t^{ji}\delta$, where $r_t^i$ is the original reward from environment without punishment. Policy $\pi^i_{\theta^i}(a^i_t|o^i_t)$ is trained to maximize
$\mathbb{E}_{\pi^i}\!\left[\sum_{t=0}^{H} \gamma^t r^{i,\text{tot}}_t\right]$
using the TD-error, where $H$ denotes the length of an episode.

\begin{figure}[t]
  \centering 
\hspace{-3em}
    \subfigure[Learning Curves]{ 
        \includegraphics[width=0.5\linewidth]{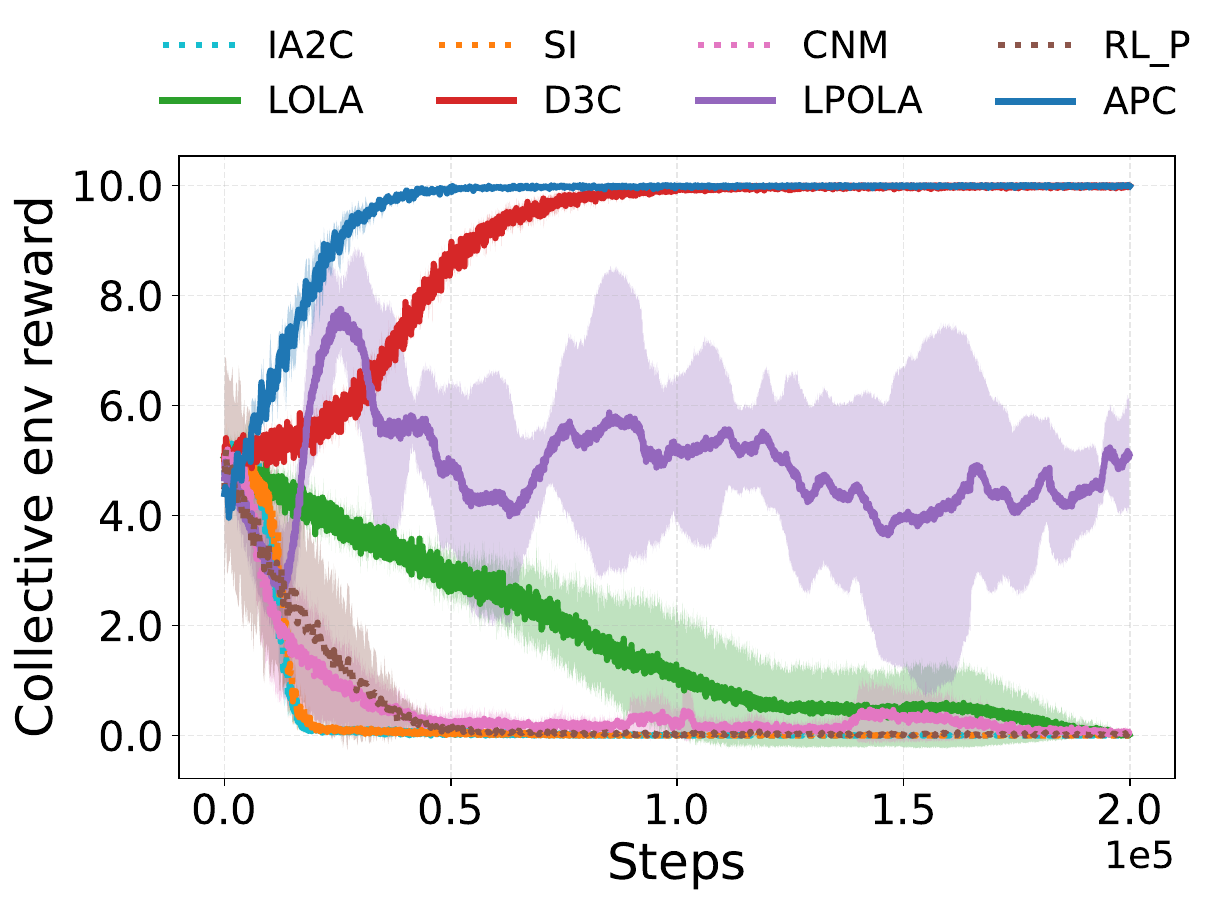}
        \label{curve}
    } 
    \hspace{-1em}
    \subfigure[Parameter Analysis]{ 
        \includegraphics[width=0.5\linewidth]{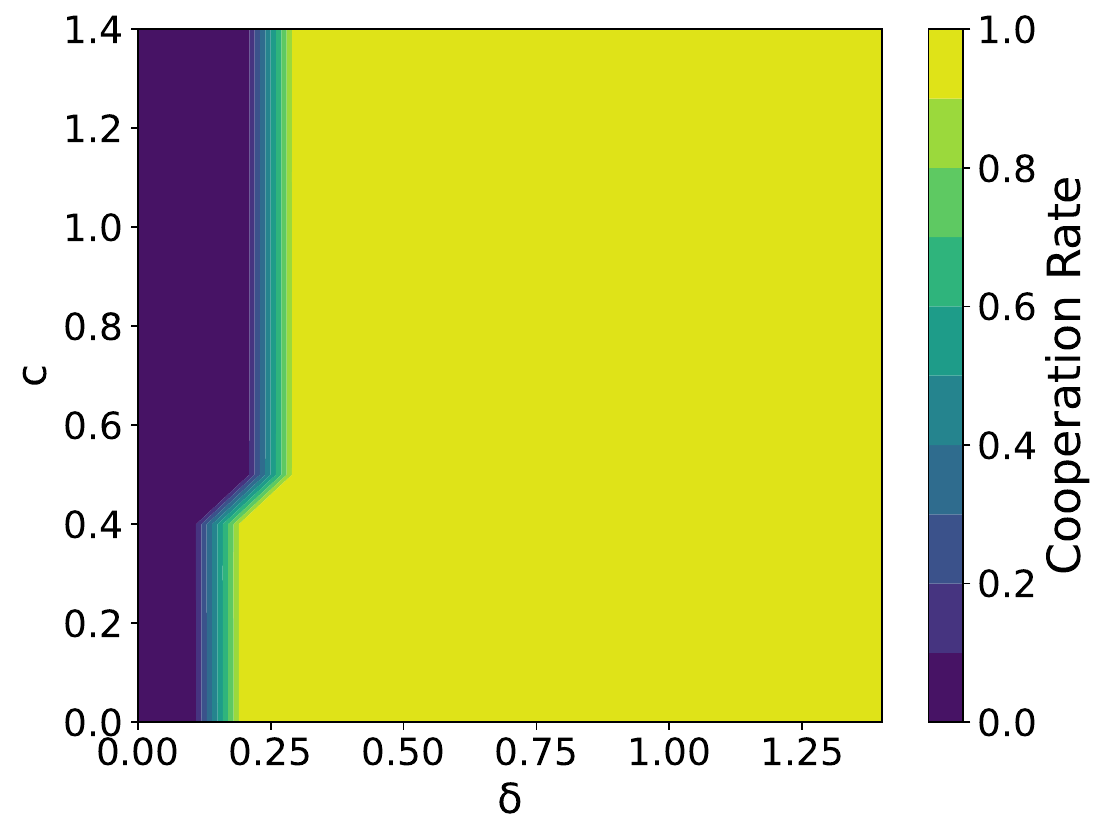}
        \label{cost_fine}
    } 
    \hspace{-3em}
\caption{(a) Learning curves for IPGG of self-play training. (b) Cooperation rate of \method after convergence across different combinations of $c$ and $\delta$ in IPGG. The X-axis and Y-axis represent the values of $\delta$ and $c$, respectively, where $\delta, c \in [0:0.1:1.4]$.}
\end{figure}

\begin{figure*}[t]
  \centering 
  \hspace{-1em}
  \subfigure[Coingame]{ 
    \includegraphics[width=0.24\linewidth]{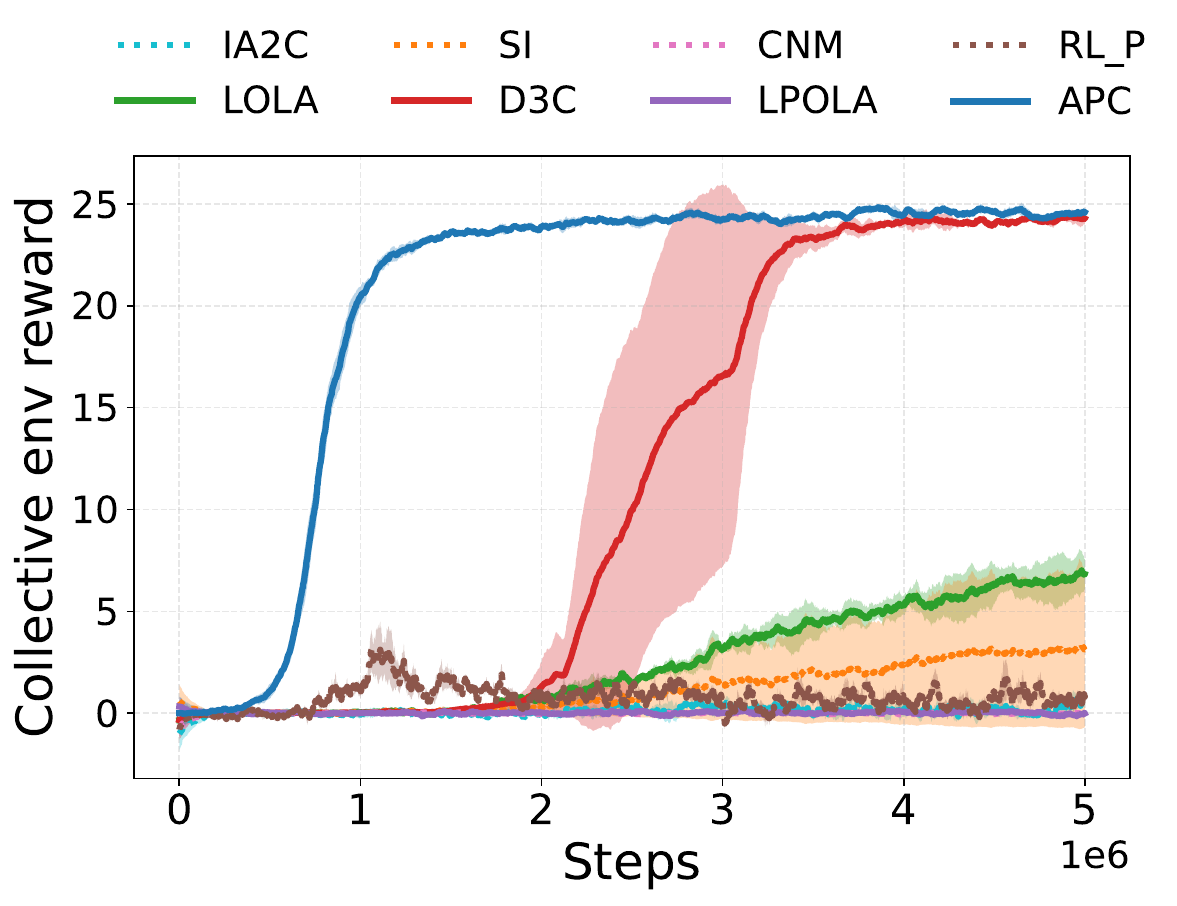}
    \label{Coingame}
  } 
  \hspace{-1em}
  \subfigure[SSG]{ 
    \includegraphics[width=0.24\linewidth]{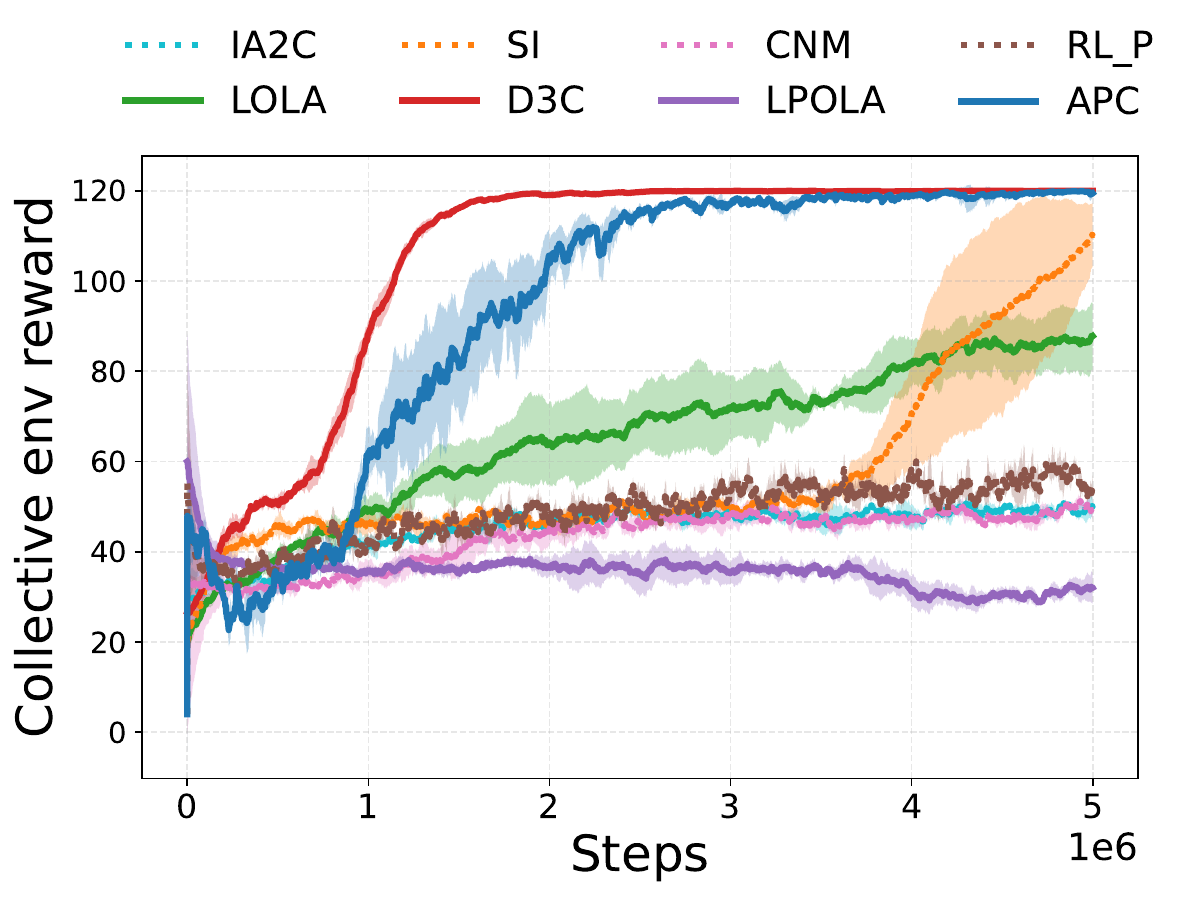}
    \label{SSG}
  } 
  \subfigure[SSH]{ 
    \includegraphics[width=0.24\linewidth]{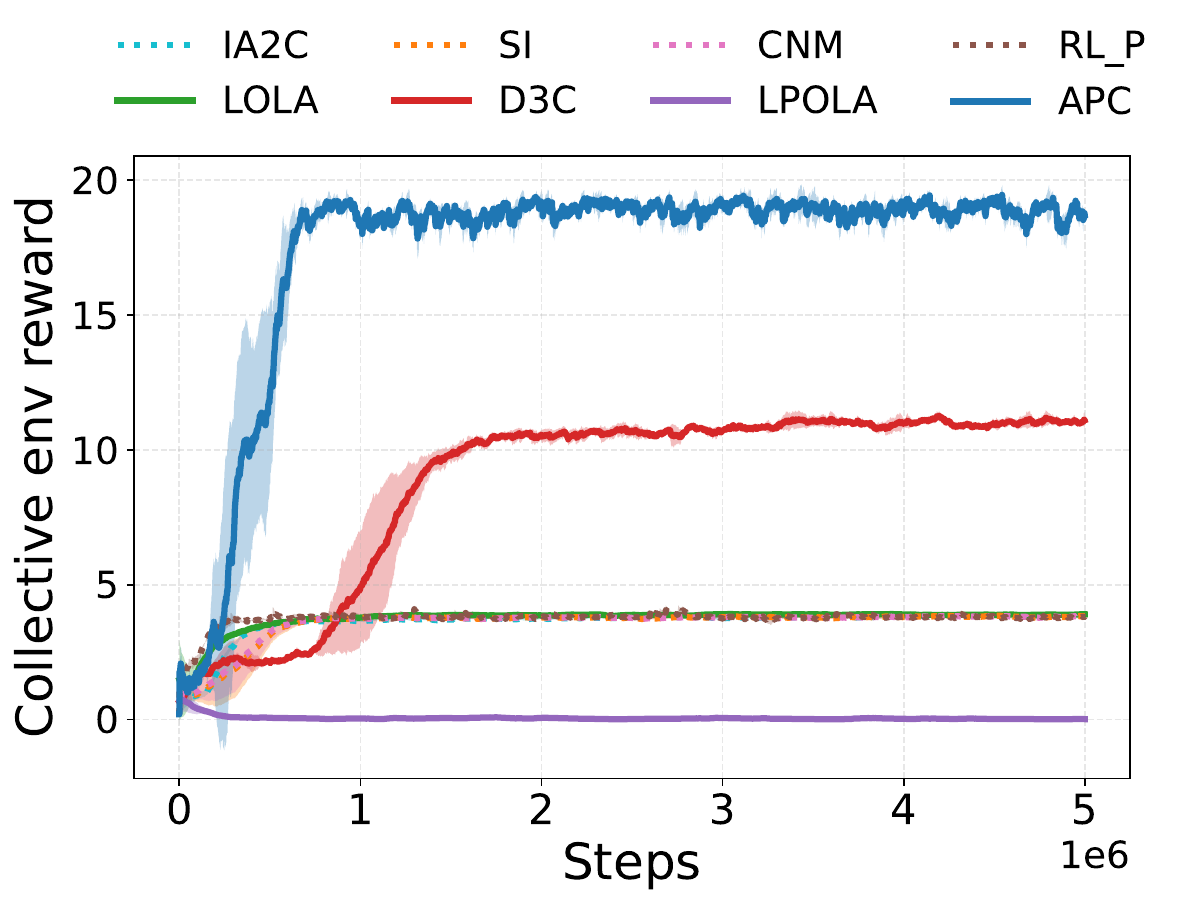}
    \label{SSH}
  } 
  \subfigure[Foraging]{ 
    \includegraphics[width=0.24\linewidth]{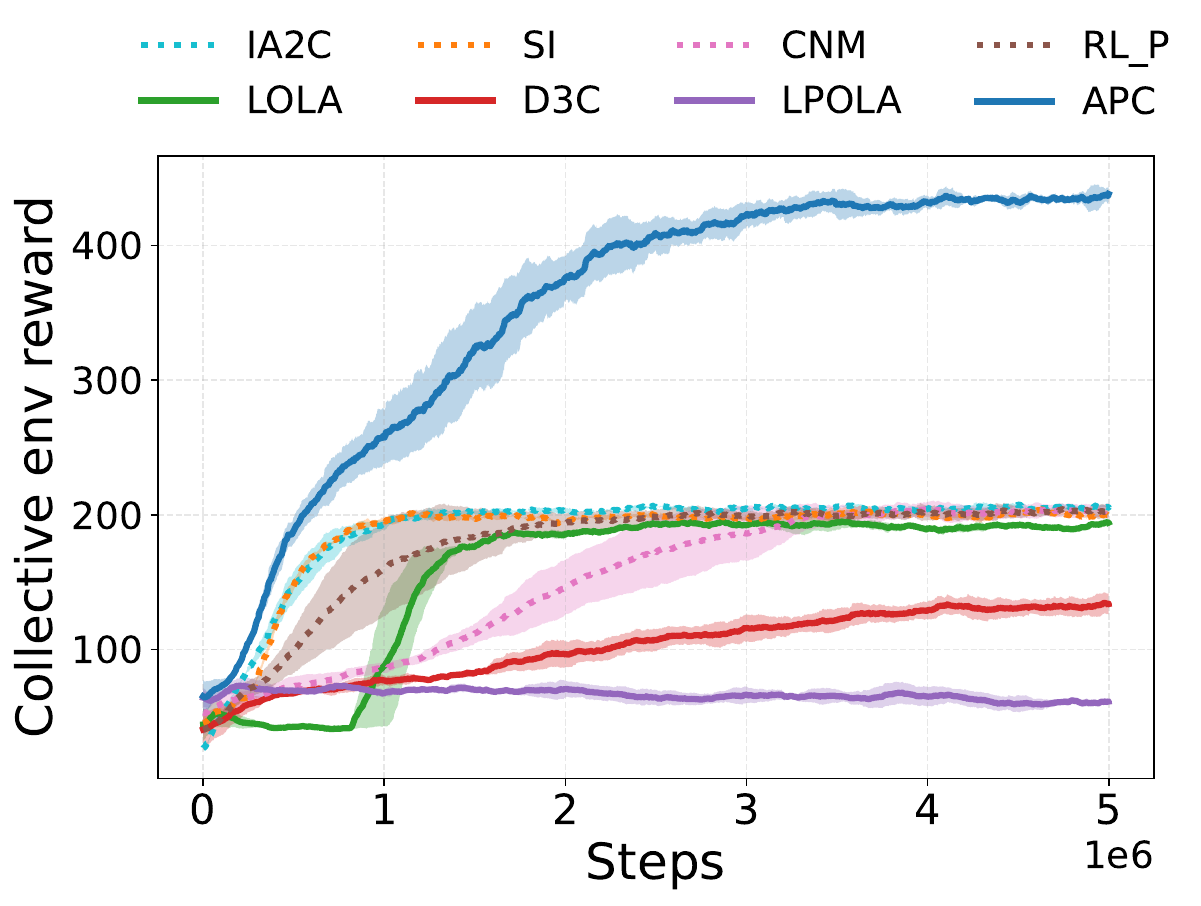}
    \label{Foraging}
  } 
  \hspace{-1em}
  \caption{Learning curves for four types of SSDs of self-play training. The curves represent the collective environment rewards, with all data collected from five training runs using different random seeds. The solid lines indicate the mean across runs, and the shaded areas represent the standard deviation. Note that all experimental results in this paper are evaluated over five random seeds, and this information will be omitted in the subsequent figure captions for brevity.} 
  \label{fig:training_curve}
\end{figure*}

\begin{figure*}[t]
  \centering 
\hspace{-1em}
    \subfigure[IPGG-APr]{ 
        \includegraphics[width=0.24\linewidth]{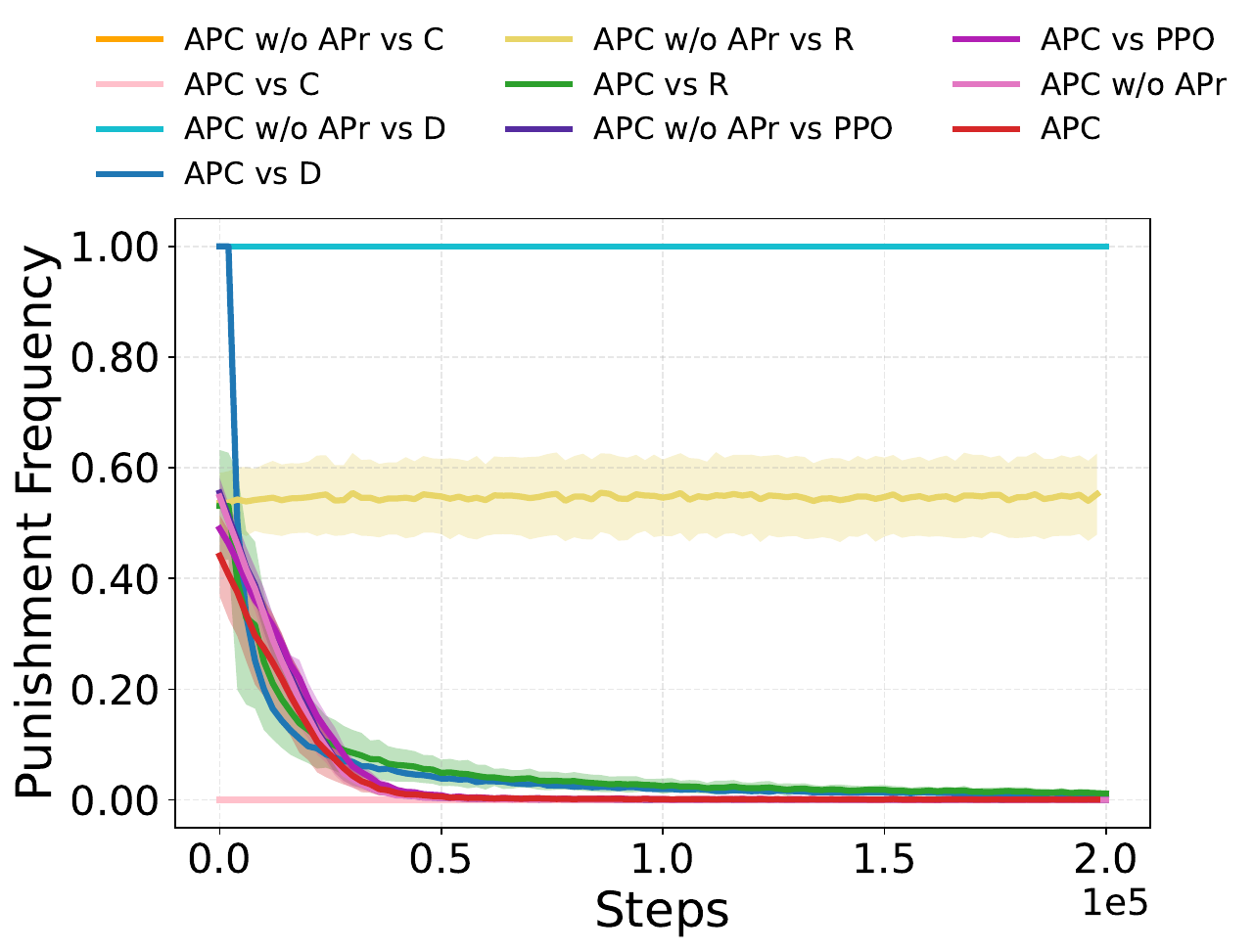}
        \label{pgg_punish_rate}
    } 
    \hspace{-1em}
    \subfigure[MIPGG-AIn]{ 
        \includegraphics[width=0.24\linewidth]{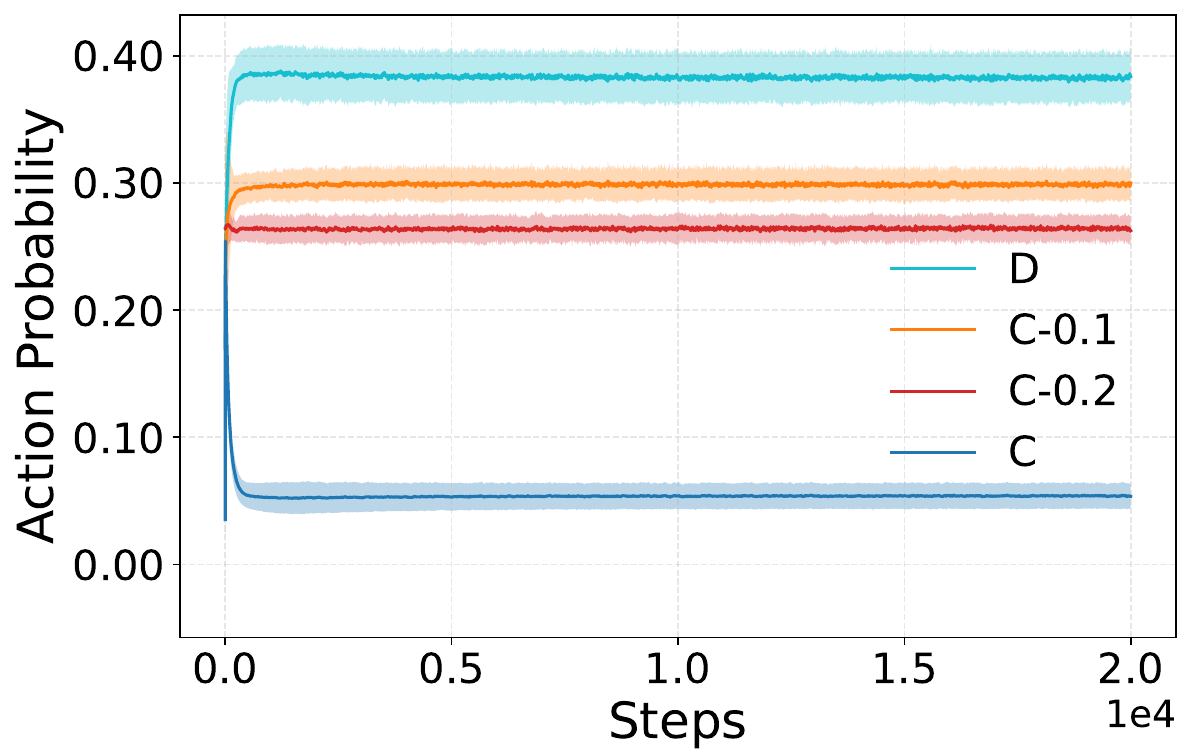}
        \label{pgg_probability}
    } 
    \subfigure[SSH-APr]{ 
        \includegraphics[width=0.24\linewidth]{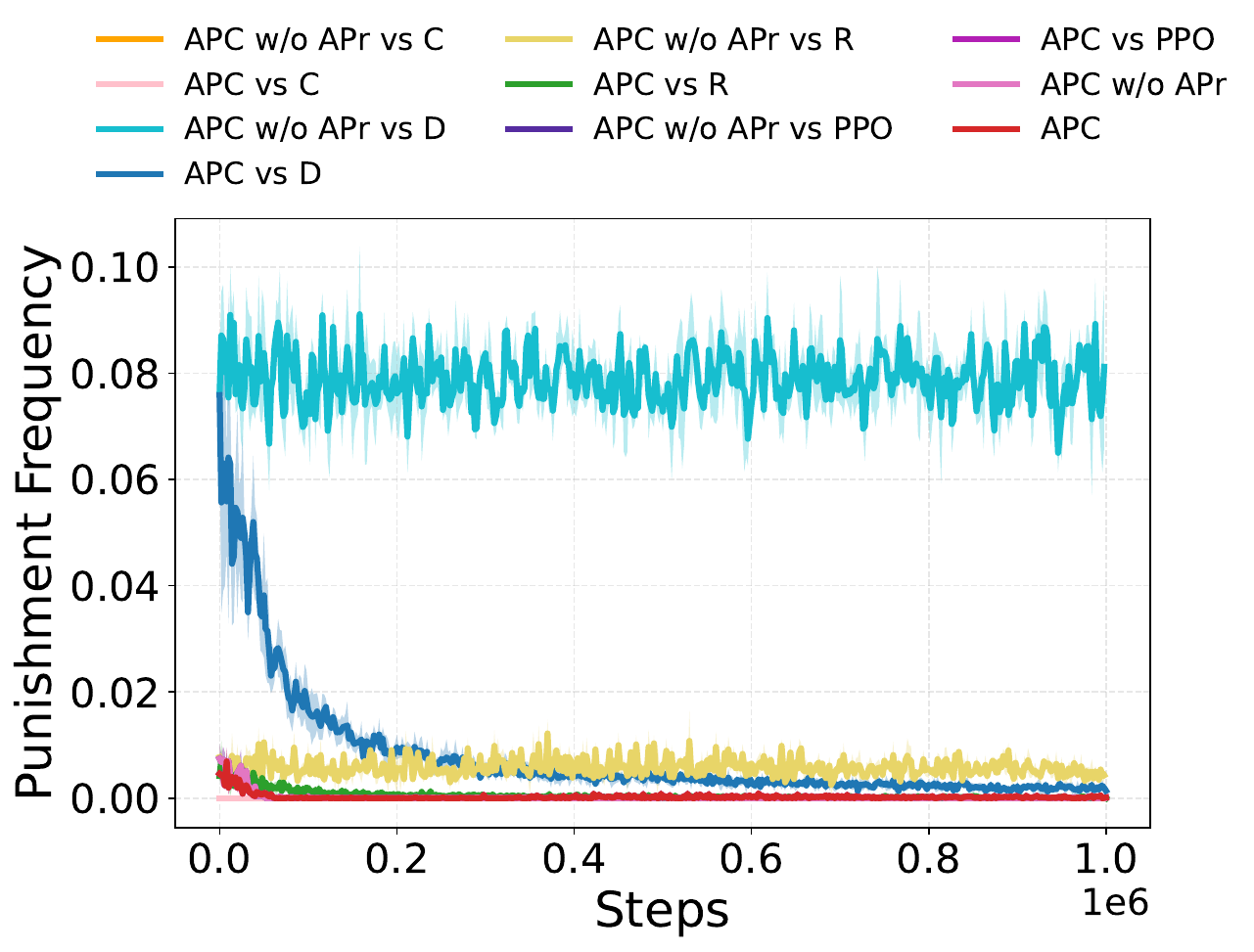}
        \label{ssh_punish_rate}
    } 
    \hspace{-1em}
    \subfigure[MSSH-AIn]{ 
        \includegraphics[width=0.24\linewidth]{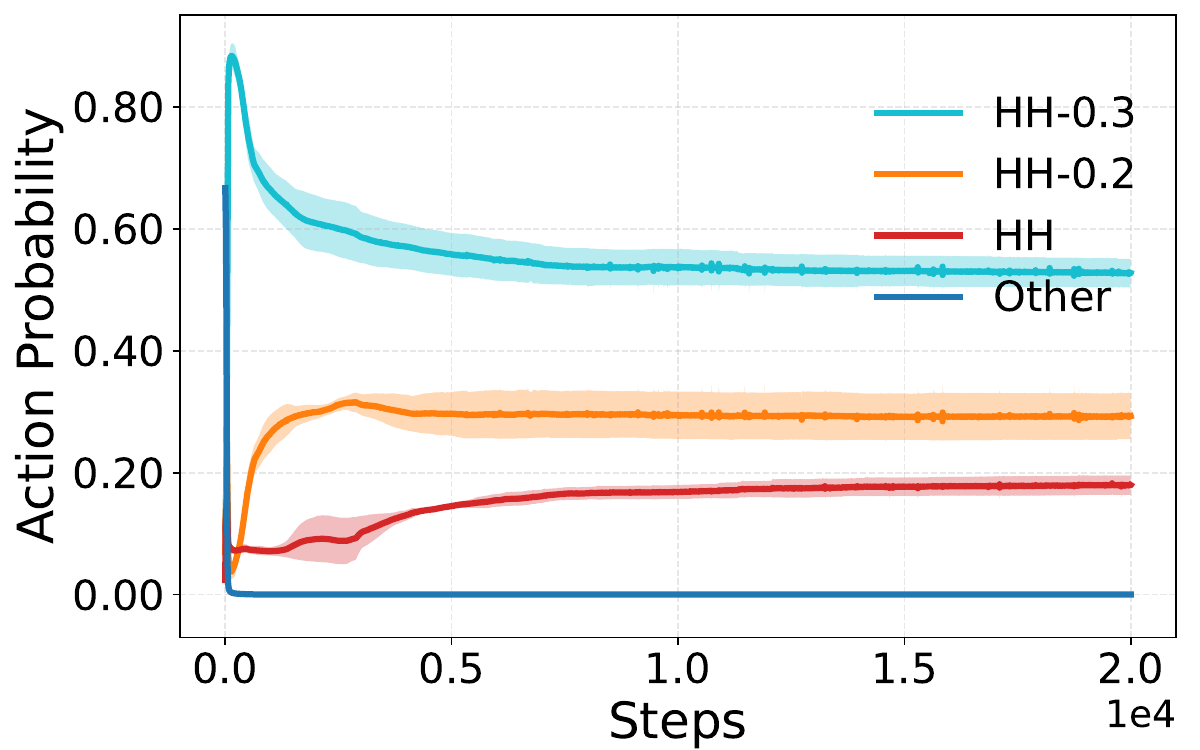}
        \label{ssh_probability}
    } 
    \hspace{-1em}
\caption{Adaptive punishment capability of APC. (a,c) APC adjusts punishment frequency: high for defection, low for cooperation, and decreases when punishment proves ineffective. (b,d) Learning curves of the action probability distribution of DPN in MIPGG and MSSH. APC scales Adaptive Intensity (AIn): stronger for severer defections, achieving proportional response.}
\label{fig:ada_punish}
\end{figure*}

\subsection{Analysis in Iterated Public Goods Game}

In Iterated Public Goods Game (IPGG), $n$ agents each decide whether to contribute an endowment $e$ to a common pool, which is multiplied by $r$ and split equally.

\textbf{Proposition 1.} \method agents converge to mutual cooperation in Iterated Public Goods Game.


\textbf{Proof.} Let $b$ the expected punishment cost. Given the probability of others cooperating $p_c$, the expected rewards for contribution (cooperation, $r^i_c$) and no contribution (defection, $r^i_d$) are formulated as:
\begin{align}
    r^i_c &= p_c \frac{(n-1)er}{n} + \frac{er}{n} - e - b \\
    r^i_d &= p_c \frac{(n-1)er}{n} - (n-1)\delta - b
\end{align}
The contribution probability $\theta^i$ is updated via gradient ascent:
\begin{equation}
\begin{aligned}
    {\theta}^i &= \theta^i + \alpha \nabla_{\theta^i} V^i(\theta) \\
    &\approx \theta^i + \frac{\alpha}{1 - \gamma} \cdot \nabla_{\theta^i} \left[ \theta^i \cdot r^i_c + (1 - \theta^i) \cdot r^i_d \right] \\
    &= \theta^i + \frac{\alpha}{1-\gamma} \left( \frac{er}{n} - e + (n-1)\delta \right)
\end{aligned}
\end{equation}

The parameters are typically set to $[n, e, r] = [5, 1, 3]$, and the gradient becomes positive when $\delta > 0.1$. As shown in Fig. \ref{curve}, experiments on the IPGG confirm that APC effectively promotes cooperation. Sensitivity analysis of punishment parameters (see Fig. \ref{cost_fine}) reveals that full cooperation is achieved when $\delta \geq 0.3$ for $c \in [0, 1.4]$, and when $\delta = 0.2$ for $c \in [0, 0.4]$. Experimental results are highly consistent with our theoretical derivations. Specifically, the failure of cooperation at $\delta = 0.2$ for $c \in [0.5, 1.4]$ can be attributed to the fact that a small $\delta$ exerts insufficient deterrence, while a high cost $c$ simultaneously hinders the learning of cooperative policies. However, as $\delta$ increases, this negative impact of $c$ is effectively diminished.

\section{Experiment}
\subsection{Environments}
\label{sec:env}

We evaluate \method on four spatiotemporal-extended SSDs mixed-motive games. The \textbf{Coin Game} \cite{coingame} ($5\times5$ map, 2 agents) is a spatial extension of the Iterated Prisoner's Dilemma: agents receive 1 point for any coin but penalize partners by 2 points when collecting the opponent's color. In the \textbf{Sequential Snowdrift Game (SSG)} ($8\times8$ map, 4 agents), clearing a snow pile grants 6 points to all agents while the clearing agent incurs a cost of 4, incentivizing free-riding. \textbf{Sequential Stag-Hunt (SSH)} \cite{GO} ($8\times8$ map, 4 agents) requires coordination, as hunting a stag yields 10 shared points but requires at least two participants, whereas hunting a hare is a low-reward (1 point) individual task. Finally, \textbf{Foraging} \cite{assum3} ($10\times10$ map, 12 agents) involves common agents collecting berries (3 points) and special agents who can harvest forbidden berries (4 points); the latter triggers permanent resource degradation, reducing common berry values to 1 point. To deter defection, \method introduces punishment parameters $[\delta, c]$, representing the fine imposed and the cost incurred by the punisher. We set $c = \delta$, choosing $\delta$ to offset potential gains from defection. Parameter configurations $[\delta, c]$ are: $[0.7, 0.7]$ for IPGG, $[1.1, 1.1]$ for Coin Game, $[2.1, 2.1]$ for SSG, $[0.4, 0.4]$ for SSH, and $[0.7, 0.7]$ for Foraging.


\subsection{Implementations}
\label{sec:imple}
We adopt a Decentralized Training Decentralized Execution (DTDE) architecture with independently parameterized networks. The defection predictor consists of two convolutional layers for encoding, an LSTM for temporal modeling, and ReLU-activated fully connected layers, processing multi-channel binary tensors. The policy network follows an actor-critic framework, where both components share an architecture similar to the predictor. The actor's output dimension matches the predictor's, while the critic produces a single scalar value.


\subsection{Baselines}
Independent Advantage Actor-Critic (IA2C) is a classic gradient-based reinforcement learning algorithm suitable for agents to learn policies under completely independent conditions \cite{a2c}. LOLA considers the learning process of other agents when updating its own policy parameters \cite{oppo1}. SI achieves coordination by rewarding agents for having causal influence over other agents’ actions \cite{intrinc3}. D3C guides self-interested agents toward collectively efficient cooperative equilibria by having them mix rewards and follow the gradient of an efficiency bound during learning \cite{d3c}. To enable punitive capability, all above methods are equipped with a punitive action within their action spaces. We additionally introduce three punishment-based methods for comparison. LPOLA simultaneously predicts environmental actions and determines which actions of other agents to penalize, applying penalties when predictions match reality \cite{rl_punish1}. CNM fosters cooperation by establishing social norms, whose punishments are socially enforced based on group consensus \cite{rl_punish2}. The core idea of both LPOLA and CNM methods is to encourage punishment by attaching pseudo-rewards. RL Punish utilizes two distinct networks: a policy network for making primary action decisions and a punishment network for deciding which agents to penalize. Both networks are trained via the Advantage Actor-Critic (A2C) algorithm with the objective of maximizing the collective environmental reward.

\begin{table}[t]
\centering
\caption{Collective Environmental Reward of Ablation Study}
\label{tab:win_rates}
\renewcommand{\arraystretch}{1.2} 
\begin{tabular}{l|cccc}
\toprule
\textbf{Collective Env Reward} & \textbf{Coingame} & \textbf{SSG} & \textbf{SSH} & \textbf{Foraging} \\ 
\midrule
\textbf{APC}     & 24.628 & 119.618 & 18.96 & 439.283 \\
\textbf{APC w/o DPN} & 0.010 & 31.804 & 0.014 & 60.645 \\
\textbf{APC w/o APr}& 24.617 & 119.458 & 18.959 & 438.760 \\
\bottomrule
\end{tabular}
\end{table}

\subsection{Main Results}


In Coingame, \method most rapidly guides agents to avoid collecting coins of other agents' colors, significantly improving collective rewards (Fig. \ref{Coingame}), whereas D3C achieves this at a relatively slower pace. Although LOLA and SI learn to some extent to prevent agents from picking other agents' coins, both fail to completely avoid such behavior. For CNM, both agents eventually converge on punishing all actions, which leads to solely optimizing pseudo-rewards and results in very low total return. Consequently, none of the agents successfully learn to collect coins. In addition, agents of other baselines fail to learn an effective punishment strategy and fail to realize that picking coins of other agent’s color would lead to punishment. As a result, they continue to selfishly collect coins.

In SSG, \method agents learn to clear all snow piles faster than D3C, SI, and LOLA, achieving near-optimal collective rewards (Fig. \ref{SSG}). However, LPOLA, due to interference from pseudo-rewards associated with punitive actions, neglects the pursuit of actual environmental rewards, falling below IA2C. Other baselines remain trapped in the snowdrift dilemma by expecting others to clear snow while free-riding, leaving numerous piles uncleared.

In SSH, agents in \method, concerned about being punished by other agents for hunting hares, ultimately choose to cooperate in hunting stags, thereby maximizing collective rewards (Fig.~\ref{SSH}). D3C gets it into the lazy problem \cite{VDN}, and early hunting leads to moving out of the environment and failing to obtain the group rewards of others’ later hunting. Thus, D3C may not hunt until the last few steps, and likely miss the the opportunity of cooperating to hunt stags. Due to pseudo-rewards, LPOLA also neglects the pursuit of actual environmental rewards, resulting in performance inferior to IA2C. In other baseline methods, agents, seeing only short-term personal rewards, choose to hunt hares, failing to engage in group cooperation. 


In Foraging, \method successfully prevents special agents from collecting forbidden berries, thereby enhancing the subsequent collective returns and achieving optimal performance (as shown in Fig. \ref{Foraging}). In contrast, although D3C performed well in IPGG and the first three SSDs, its performance in Foraging was even worse than that of IA2C. Meanwhile, LPOLA was still affected by pseudo-rewards, which led to its performance being inferior to IA2C. Other baseline methods resulted in common agents consistently collecting common berries while special agents persistently pursued forbidden berries, with none learning to prevent the special agents from collecting the forbidden ones. This failure led to a significant gap between the overall reward and the optimal outcome.

\subsection{Ablation Study}
\label{sec:ablation}
To further evaluate the performance of Defection Predictor Network (DPN) and Adaptive Probability (APr) (Eq.(\ref{eq:punishment_prob})), two ablation experiments are designed. The first ablation experiment is as follows: we removed the trained DPN and replaced it with a randomly initialized policy network to verify the critical role of a reasonable defection predictor in improving overall performance. This method is referred to as \method w/o DPN. The second ablation experiment is as follows: rather than using an adaptive probability, we simply set it to 1 throughout, which is denoted as \method w/o APr.


The results of ablation experiments are shown in Table~\ref{tab:win_rates}. For \method w/o DPN, due to the failure to learn effective and reasonable punishment policy, it is unable to leverage the advantages of the punishment. The performance of \method w/o DPN was generally poor, there was no significant improvement in performance. For \method w/o APr, in the self-training setting (i.e., \method w/o APr agents interacting with each other), opponents learn to avoid punishment and reduce their defection behaviors, leading to low punishment frequency, despite the agent maintaining a high probability $p^{ij}$. As a result, its performance remains close to that of \method. However, when facing rule-based opponents (see Fig.~\ref{pgg_punish_rate}, \ref{ssh_punish_rate}), \method w/o APr agent sustains high punishment frequency, failing to adapt. This highlights the importance of adaptive probability in avoiding ineffective and excessive punishment when opponents do not respond to it. It is worth noting that in the results of APC w/o APr vs Defect in Fig.~\ref{ssh_punish_rate}, the noise is caused by the stochastic nature of SSH, where the number of hare-hunting actions (defections) may vary across different time windows.

\subsection{Adaptive Punishment Capability}

We evaluate APC's adaptive capability from two perspectives: \textbf{frequency} and \textbf{intensity}. 

First, we assess \textit{adaptive punishment frequency} by examining whether APC dynamically adjusts sanctions based on opponent behavior. The focal APC agent was tested against rule-based (Always-Defect, Always-Cooperate, Random) and shapeable (PPO, APC) opponents in IPGG and SSH. As shown in Fig.~\ref{pgg_punish_rate} and \ref{ssh_punish_rate}, APC initially targets defectors and random agents with high-frequency punishment while sparing cooperators. Over time, punishment self-regulates: the agent reduces $p_t^{ij}$ (Eq.(\ref{eq:punishment_prob})) if sanctions prove ineffective or if opponents (PPO/APC) shift toward cooperation, thereby avoiding excessive costs while maintaining behavioral control.

Second, we evaluate \textit{adaptive punishment intensity} using modified environments (MIPGG and MSSH) featuring varying defection severities. In MIPGG, we introduced partial contributions (C-0.2, C-0.1) alongside full contribution (C) and defection (D). In MSSH, actions included hunting hare with additional payoff reductions (HH-0.2, HH-0.3) compared to standard hare hunting (HH) or other cooperative actions. Results in Fig.~\ref{pgg_probability} and \ref{ssh_probability} show APC's DPN assigns proportionate Punishment Intensity Weights (PIW). Specifically, PIW followed Eq.(\ref{eq:punishment_intensity}), yielding $[1, 0.79, 0.70, 0]$ for [D, C-0.1, C-0.2, C] in MIPGG, and $[1, 0.543, 0.360, 0]$ for [HH-0.3, HH-0.2, HH, Other] in MSSH. These findings confirm APC's ability to impose sanctions proportional to defection severity.

\section{Conclusions}
We propose Adaptive Punishment for Cooperation (\method), which combines dynamic punishment intensity with explicit defection awareness to align punishment intensity with defection severity, reducing exploitation and defection frequency. A limitation of this work is that \method has only been evaluated in relatively simplified environments. Its effectiveness and scalability in more complex and realistic multi-agent scenarios remain to be validated. Additionally, this work focuses solely on punishment; how \method complements reward-based method for cooperation remains an open question.

\bibliographystyle{IEEEtran} 
\bibliography{iclr2026_conference}    

\end{document}